\documentclass[prd,nofootinbib,preprintnumbers,floatfix]{revtex4}  
\usepackage[plainpages=false, colorlinks=true, anchorcolor=blue, linkcolor=blue, citecolor=blue, bookmarks=false]{hyperref}

\usepackage{amsmath}
\usepackage{graphicx}
\begin{document}
\newcommand{\apjl}{Astrophys. J. Lett.}
\newcommand{\apjs}{Astrophys. J. Suppl. Ser.}
\newcommand{\aap}{Astron. \& Astrophys.}
\newcommand{\aj}{Astron. J.}
\newcommand{\pasp}{PASP}
\newcommand{\araa}{Ann. Rev. Astron. Astrophys. } 
\newcommand{\aapr}{Astronomy and Astrophysics Review}
\newcommand{\ssr}{Space Science Reviews}
\newcommand{\mnras}{Mon. Not. R. Astron. Soc.}
\newcommand{\apss} {Astrophys. and Space Science}
\newcommand{\jcap}{JCAP}
\newcommand{\na}{New Astronomy}
\newcommand{\pasj}{PASJ}
\newcommand{\pasa}{Pub. Astro. Soc. Aust.}
\newcommand{\physrep}{Physics Reports}
\newcommand{\dnu} {$\left(\frac{\Delta \nu}{\nu}\right)$}
\title{Classification of Pulsar Glitch Amplitudes using Extreme Deconvolution }
\author{Swetha  Arumugam$^{1}$}%
 \email{ee20btech11005@iith.ac.in}

\author{Shantanu Desai$^{2}$ }%
 \email{shntn05@gmail.com}

\affiliation{$^{1}$ Department of Electrical Engineering, IIT Hyderabad, Kandi, Telangana, India 502284}

\affiliation{$^{2}$ Department of Physics, IIT Hyderabad, Kandi, Telangana, India 502284}
\begin{abstract}
We carry out a classification of the glitch amplitudes of radio pulsars  using Extreme Deconvolution technique  based on the  Gaussian Mixture Model, where the observed uncertainties in the glitch amplitudes \dnu are taken into account. Our dataset consists of  699 glitches from 238 pulsars.
We then use information theory criteria such as AIC and BIC to determine the optimum number of glitch classes. We find that both AIC and BIC  show   that the pulsar glitch amplitudes can be optimally described  using a bimodal distribution.  The mean values of \dnu for  the two components are equal to $4.79 \times 10^{-9}$ and $1.28 \times 10^{-6}$, respectively with standard deviation given by 1.01 and 0.55 dex. We also applied this method to classify the pulsar inter-glitch time intervals, and we find that AIC prefers two components, whereas BIC prefers a single component. The unified data set and analyses codes used in this work have been made publicly available.
\end{abstract}
\maketitle
\section{Introduction}
Pulsars are rapidly rotating neutron stars which emit pulsed radio emission~\cite{Kaspi}. Although the periods of most pulsars are stable, a small fraction of young pulsars show abrupt discontinuities in their period. Glitches constitute one such discontinuity, where there is a  sudden abrupt increase in the rotational frequency, with frequency jumps ranging from $(10^{-4}- 10^2)\mu$ Hz~\cite{Espinoza,Manchester}. The glitch is followed by a recovery stage in which the rotation frequency asymptotes to the pre-glitch behaviour. The   glitch amplitude is usually parameterized by \dnu, where $\Delta \nu$ is the change in rotational frequency and $\nu$ is the observed frequency. This amplitude ranges from $10^{-11}- 10^{-4}$, whereas  the inter-glitch time interval varies between 20-1000 days~\cite{Eya2018}.
Ever since the detection of the first glitch in the Vela pulsar in 1969~\cite{Radh69,Downs}, glitches have proved to be a wonderful laboratory for probing the physics of the  neutron star, in particular   the nature of superfluid in the interior of the neutron star~\cite{Baym}. However, the full details  of the glitch mechanisms are still not completely understood~\cite{Melatos}. Therefore,  there are large number of observational campaigns on current as well as future radio telescopes to get more insights into the nature of glitches and their causes~\cite{Singha21}. The study of pulsar glitches will also be one of the flagship science goals carried out during the SKA era~\cite{Singha22}.
Such glitches  have also been observed in millisecond pulsars~\cite{Cognard,Mckee} and also magnetars~\cite{Dib}, where  anti-glitches (corresponding to spin-down  in the neutron star period)   have been detected~\cite{Archibald}.

The study of pulsar glitches  is very important for gravitational wave astronomy. Pulsar glitches could result in a gravitational wave burst~\cite{Melatos08}. Studying the statistical characteristics of these glitches is therefore important for designing search pipelines to look for such gravitational wave bursts~\cite{Clark,Hayama,Prix}. Furthermore, glitches could also  affect   the sensitivity of continuous gravitational wave (CW) searches~\cite{Ashton}. It is important to understand the probability of occurrence of a glitch during a CW and the impact of a probable glitch on its detectability. The statistical characteristics of the observed glitch population has also been used  to extrapolate the glitch magnitudes for the unobserved neutron star population to understand its implications for CW searches~\cite{Ashton}.

There have been several studies which have looked at the statistical distribution of glitch amplitudes, which  have alluded to a bimodal distribution~\cite{Lyne2000,Wang2000,Espinoza,Yu2012,Fuentes,Konar,Ashton,Eya2017,Eya2018,Basu2022}. A few works have previously  carried out a GMM based analysis of the glitch amplitudes~\cite{Ashton,Fuentes} or the fractional glitch amplitude~\cite{Konar},  followed by Bayesian~\cite{Ashton} or  AIC based model comparison test~\cite{Fuentes}.
However none of these works have included the associated errors  in the fractional glitch amplitudes. In addition to the aforementioned importance of understanding the statistical nature of glitches for gravitational wave astronomy, the bimodal nature  of the glitch amplitude distribution is important for constructing unified models of the glitch phenomenon~\cite{Celora}. Therefore it is important to re-assess the robustness of the claim of bimodality  of the glitch amplitude distribution, when the uncertainties are incorporated, which hitherto has never been done.

For this purpose, we carry out a clustering based analysis of pulsar glitch amplitudes by applying the Extreme Deconvolution technique based on the Gaussian Mixture model~\cite{Bovy} and  determine the optimum number of classes using information theoretical criteria~\cite{Krishak}. This work  follows up on our previous studies, where we applied these same techniques for the classification of pulsars (using their period and period derivatives) ~\cite{Reddy} and Gamma-Ray Bursts (using durations and hardness)~\cite{Bhave}.

This manuscript is structured as follows. The data set used for our analysis is described in Sect.~\ref{sec:dataset}. The analysis methodology is described in Sect.~\ref{sec:meth}.  Our results are discussed in Sect.~\ref{sec:results} and  we conclude in Sect.~\ref{sec:conclusions}.

\section{Data Set}
\label{sec:dataset}
A complete catalog of pulsar glitches has been independently collated in two different locations: one at Jodrell Bank (JBO) ~\cite{Espinoza}\footnote{\url{https://www.jb.man.ac.uk/pulsar/glitches/gTable.html}} and also by ATNF~\cite{ATNF}\footnote{\url{https://www.atnf.csiro.au/people/pulsar/psrcat/glitchTbl.html}}. The JBO catalog contains 664 glitches from 207 distinct pulsars. Each JBO glitch entry consists of pulsar name, glitch MJD,  glitch amplitude  and its error, the first derivative of the glitch amplitude along with its error. In this catalog, two glitches did not have any associated amplitude and error, whereas 16 glitches had no associated uncertainties for \dnu. The ATNF catalog contains 641 glitches from 211 distinct pulsars and 18 glitches without any associated pulsar. This catalog also  contains 54 entries from 32 distinct pulsars, which are not in the JBO catalog. 

For our analysis we primarily used the JBO catalogue. However, we appended to  this catalog the 54 entries from the ATNF list which were absent from JBO. We also checked to see if any of the entries with missing amplitude or error in the JBO catalog had corresponding values in the ATNF and we found 5 such events. 
After combining data from both the catalogs we had a total of 717 glitches from 238 distinct pulsars. After preprocessing the data to remove pulsars associated with negative glitch amplitude and ones with no associated uncertainties, we had a total of 699 glitches from 238 pulsars. The median value of the fractional error in the glitch amplitude is about 1.5\%.
The full dataset used for our analysis can be found at \url{https://github.com/swetha9730/XDGMM-for-Pulsar-Glitches-Classification}.
Given the large dynamic range in \dnu, we do the classification in logarithmic space using the base 10. We now describe  the methodology used for classification.

\section{Methodology}
\label{sec:meth}
\subsection{XDGMM}
Extreme Deconvolution (XDGMM) is a generalization of Gaussian Mixture Model~\citep{Feigelson17} (GMM), which incorporates the uncertainty in the observed data~\cite{Bovy,astroML,Wechsler}. It has been used for a variety of astrophysical applications, including determining the  velocity distribution from Hipparcos~\cite{Bovy}, classification of pulsars~\cite{Reddy} and GRBs~\cite{Bhave},   the three-dimensional motions  of   Sagittarius stream stars ~\citep{sagittarius_koposov},  bifurcation of  neutron star masses~\citep{Keitel}, detection of dark matter subhalo candidates~\citep{Miguel}. We provide a brief description of the XDGMM method, and use the same notation as ~\cite{Reddy,Bhave}. More details about XDGMM can be found in ~\cite{Bovy,astroML,Wechsler}. We note that GMM has also been extended to account for incomplete/missing as well as truncated data~\cite{Melchior}.

Consider a  noisy data set $x_i$, which is  related to the true values $v_i$ according to~\citep{Bovy,astroML}:
\begin{equation}
x_i=R_i v_i + \epsilon_i,
\label{eq:transform}
\end{equation}
where $R_i$ is the rotation matrix used to transform the correct values to the observed noisy dataset.   Similar to ordinary GMM, we assume that the probability density  of the true values $v$ can be written as a mixture of $K$ Gaussians given by
\begin{equation}
p(v_i) = \sum_{j=1}^K \alpha_j \mathcal{N} (v_i|\mu_j, \Sigma_j)   
\end{equation}
where $\mu_j$ and $\Sigma_j$ are the means and variances of each of the Gaussian distribution, and $\alpha_j$ is the weight of each Gaussian, subject to the constraint $\sum \limits_{j=1}^K \alpha_j=1$.
For the example in this work,  $x_i$
and $v_i$ represent the one-dimensional dataset  given by $\log$ \dnu .
We assume that the noise $\epsilon_i$ (in Eq.~\ref{eq:transform}) is a  Gaussian random variable with zero mean and variance equal to  $S_i$.  The likelihood of the model parameters ($\theta \equiv$ \{$\alpha$, $\mu$, $\Sigma$, $R_i$, $S_i$\}) for each noisy data point ($x_i$) is then given by~\citep{Bovy}:
\begin{equation}
p(x_i|\theta)=\sum_{j=1}^{K} \alpha_j \mathcal{N}(x_i|R_i\mu_j,R_i\Sigma_jR_i^{T}+S_i)
\end{equation}

The final step in XDGMM is to maximize the likelihood of the dataset with respect to the model parameters. This can be done by summing   the individual log-likelihood functions:
\begin{equation}
\underset{\theta}{\operatorname{argmax}} \;   L  = \sum_{i=1}^{N} \ln (p(x_i|\theta)),
\end{equation}
where $N$ is the total number of data points.
This objective function  is maximized using an extension of the Expectation-maximization algorithm~\citep{Bovy}.  The output of XDGMM returns a likelihood, which can then be used for   model selection.

\subsection{Model Selection}
There are basically three kinds of 
techniques used in literature to arbitrate between two models which are used to fit the data: frequentist, Bayesian, and information theoretical ones.  More details on these techniques, including pitfalls and advantages of each of them can be found in ~\cite{Liddle,Weller,Sharma} or in some of our past works~\cite{Kulkarni17,Kulkarniexoplanet,Ganguly17,KrishakDAMA,KrishakOJA,Haveesh,Krishakanais,Krishak}.
In this work, we shall use information theoretical techniques such as AIC and BIC, since they are straightforward to compute from the likelihood returned by XDGMM. We now add each of them below:
\begin{itemize}
    \item \textbf{AIC} AIC is defined by
\begin{equation}
AIC = 2p - 2 \ln L_{max} .
\label{eq:aic}
\end{equation}
\noindent where $p$ is the number of free parameters in the model and $L_{max}$ is the maximum likelihood. For our analysis, $L_{max}$ is obtained from XDGMM.
The model with the smaller value of AIC is considered as the better model and the significance can be assessed using the qualitative strength of evidence rules~\cite{Krishak}.

\item \textbf{BIC}
\begin{equation}
BIC = p \ln N - 2 \ln L_{max} .
\label{eq:BIC}
\end{equation} 
where $N$ is the number of data points and all the other terms is the same as in Eq.~\ref{eq:aic}. BIC penalizes models with additional free parameters more harshly than AIC. Similar to AIC, the model with the smaller value of BIC is considered as the favored model and the significance can be assessed using the strength of evidence rules proposed for BIC~\cite{Krishak}.

\end{itemize}

\subsection{XDGMM Implementation}
\label{sec:param}
We now apply XDGMM to our unified glitch catalog using  $\log$ \dnu  as inputs, where $\log$ refers to logarithm to the base 10. The error in this 
quantity is given by $\frac{\sigma}{2.3\Delta \nu/\nu}$, where $\sigma$ refers to the uncertainty in \dnu. We did not consider the data points for which \dnu$<0$, since we are doing the fit in logarithmic space.
We use the XDGMM implementation in the {\tt XDGMM} classe~\citep{Wechsler}, which is a wrapper to  the {\tt astroML} module. The output of  XDGMM   consists of 
the weights, means, and covariances for the input number of clusters.
In order to determine the optimum number of components, we  apply XDGMM by varying the  number of glitch classes from one to four
 and then use AIC/BIC to determine the best number among these.

\section{Results}
\label{sec:results}
\begin{figure}
\includegraphics[width=0.6\textwidth]{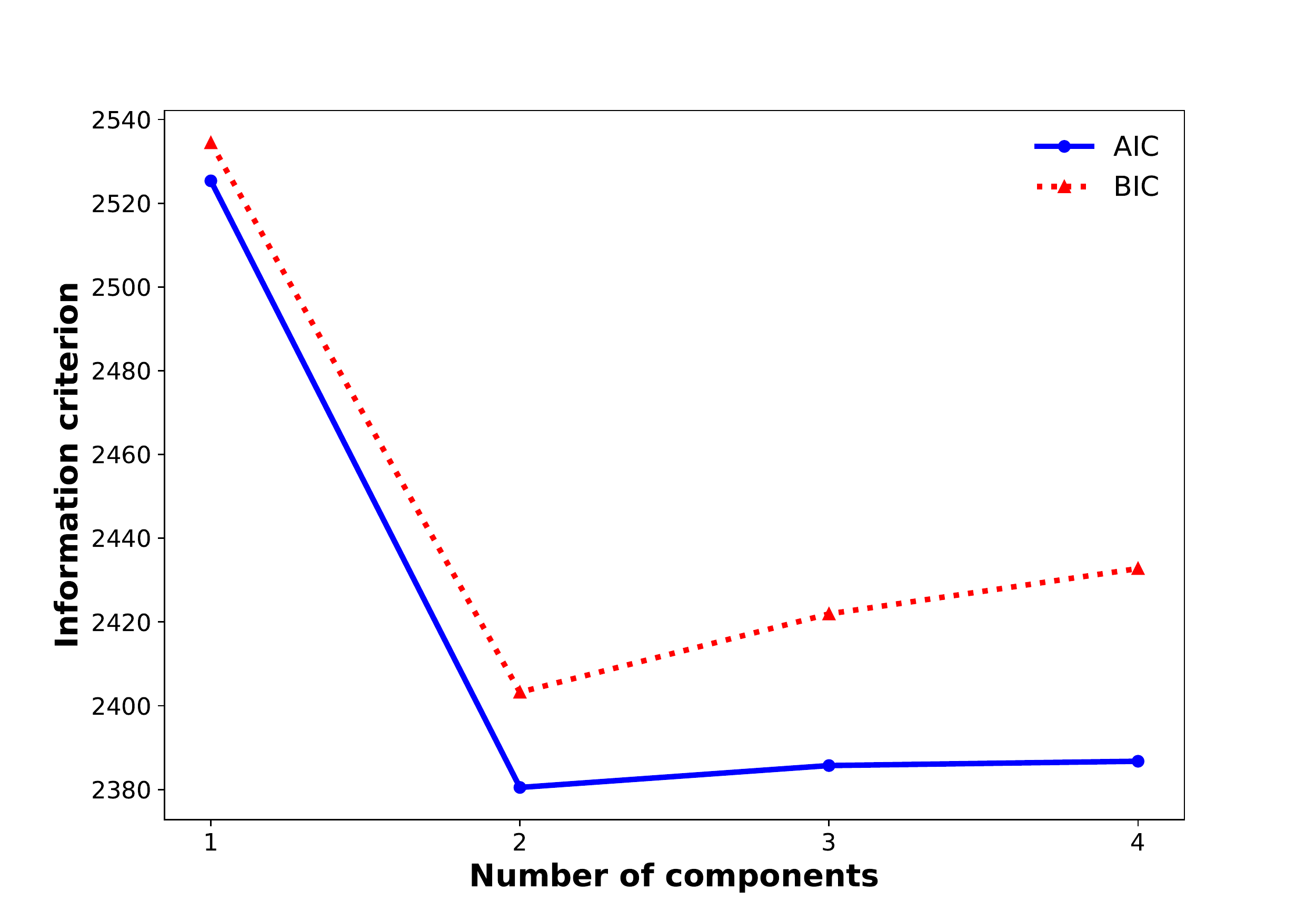}
\caption{AIC and BIC values  as a function of the number of  components. The minimum value of AIC/BIC is obtained for two components, indicating that the pulsar glitch amplitude can be adequately described using two classes.}
\label{fig:AIC}
\end{figure}

 The AIC/BIC as a function of  the number of  glitch classes can be found in Fig.~\ref{fig:AIC}. Both AIC and BIC show a minimum for two glitch classes. BIC rises sharply as we increase the number of components with ($\Delta$BIC $>10$).  Therefore, based on strength of evidence rules~\cite{Liddle,Krishak}, BIC decisively favors two classes over any other.
AIC  also shows a minimum for two components, although the  difference with respect to the third component is less than 10 ($\Delta$AIC=5). This is because  AIC is more liberal while fitting complicated models. Still both the techniques agree that the optimum number of glitch classes (based on glitch amplitude) is equal to two. 

The mean value and standard deviation of the two components ($\mu,\sigma$) are given by (-8.32,1.01) and (-5.89,0.55) respectively. This corresponds to mean fractional glitch amplitude \dnu  equal to  $4.79 \times 10^{-9}$ and $1.28 \times 10^{-6}$, respectively. There are 386 glitches which belong to the first component ($\mu_1=-8.32$) and 313 to the second component ($\mu_2=-5.89$). There are 98 glitches in the overlap region, where overlap region is defined by the region lying between $\mu_1 + 2\sigma_1$ and $\mu_2 - 2\sigma_2$.

\begin{figure}
\includegraphics[width=0.6\textwidth]{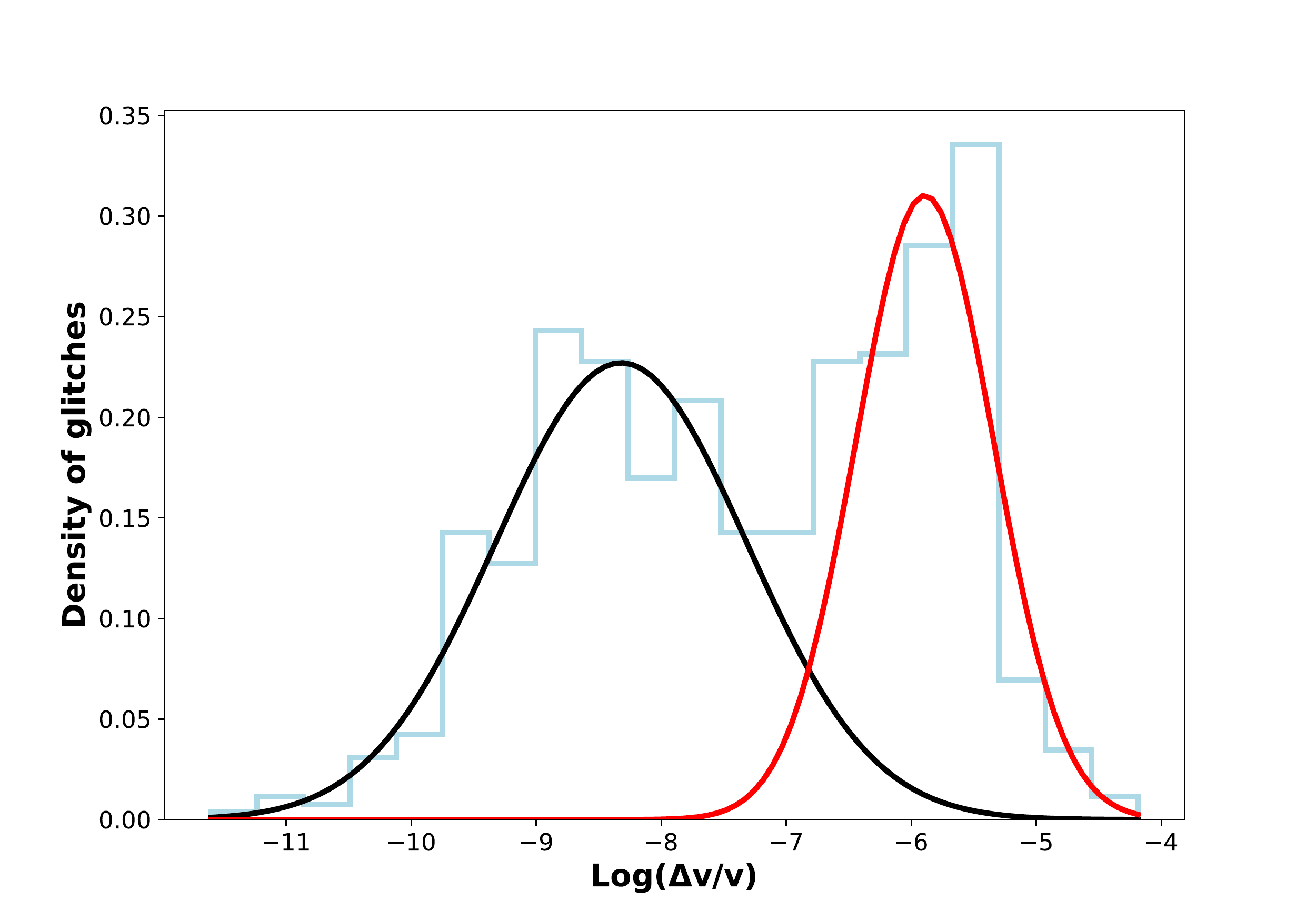}
\caption{Normalized histogram of the pulsar glitch amplitudes ($\log$ \dnu) along with the Gaussian fit obtained using XDGMM. The mean value and standard deviation for the two components ($\mu,\sigma$) are given by (-8.32, 1.01) and (-5.89, 0.55) respectively.}
\label{fig:fermi2}
\end{figure}

\subsection{Application of XDGMM to inter-glitch arrival times}
We now apply XDGMM to the pulsar inter-glitch arrival times, defined as the arrival time between multiple glitches for the same pulsars. The error in inter-glitch arrival times were obtained by adding in quadrature the uncertainty in both the time intervals. For this analysis, we used the same data-set used for XDGMM on fractional glitch amplitudes and further removed glitches which had no uncertainty in their MJD. We had a total of 658 glitches from 219 distinct pulsars and hence had 439 inter-glitch arrival times ($t_i$) (in days) after pre-processing the data. We again apply XDGMM for the glitch arrival times in logarithmic (to base 10)  space.  

\begin{figure}
\includegraphics[width=0.6\textwidth]{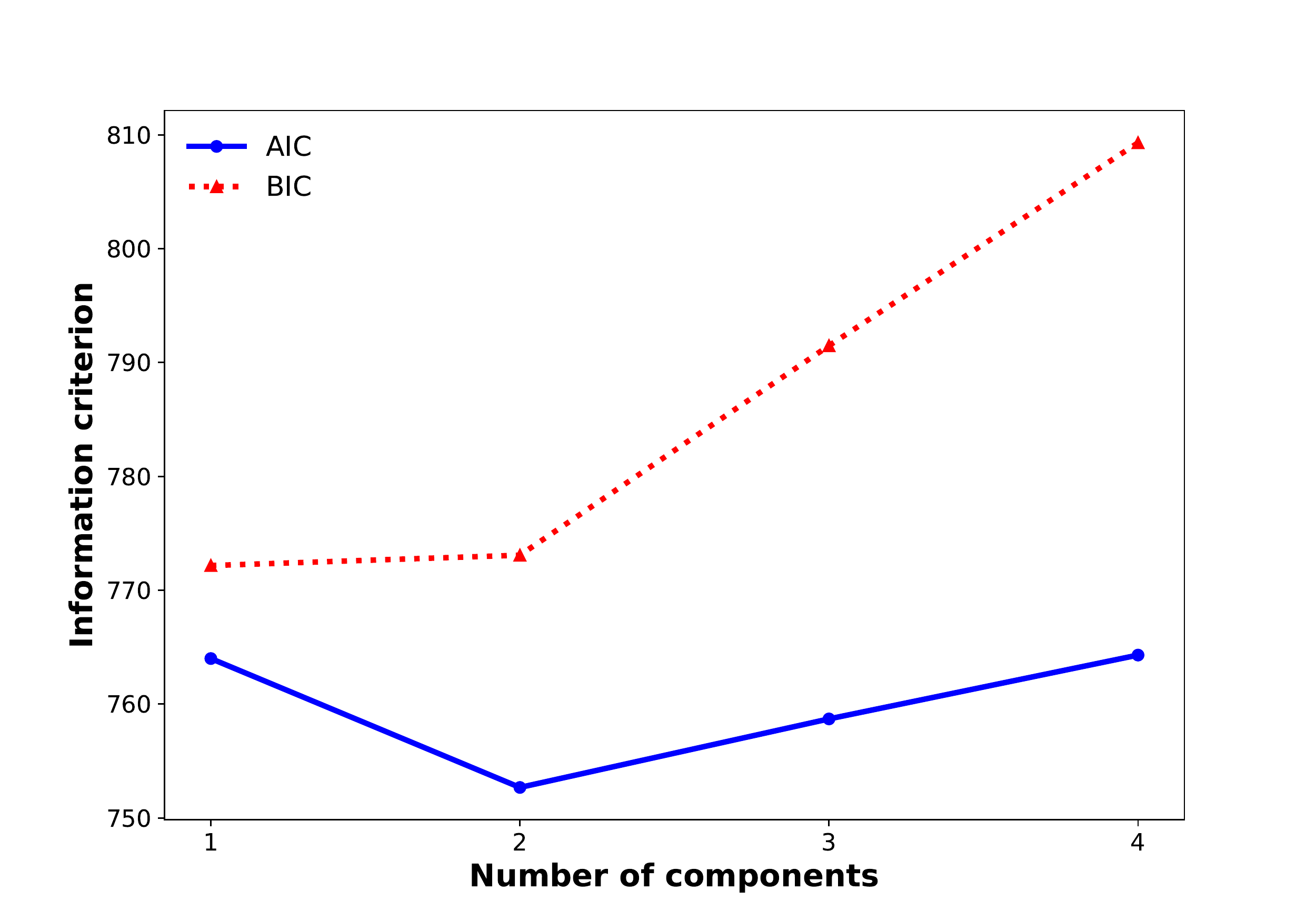}
\caption{AIC and BIC values  as a function of the number of  components. The minimum value of BIC is obtained for one components, while minimum value of AIC is obtained for two components.}
\label{fig:AIC2}
\end{figure}

\begin{figure}
\includegraphics[width=0.6\textwidth]{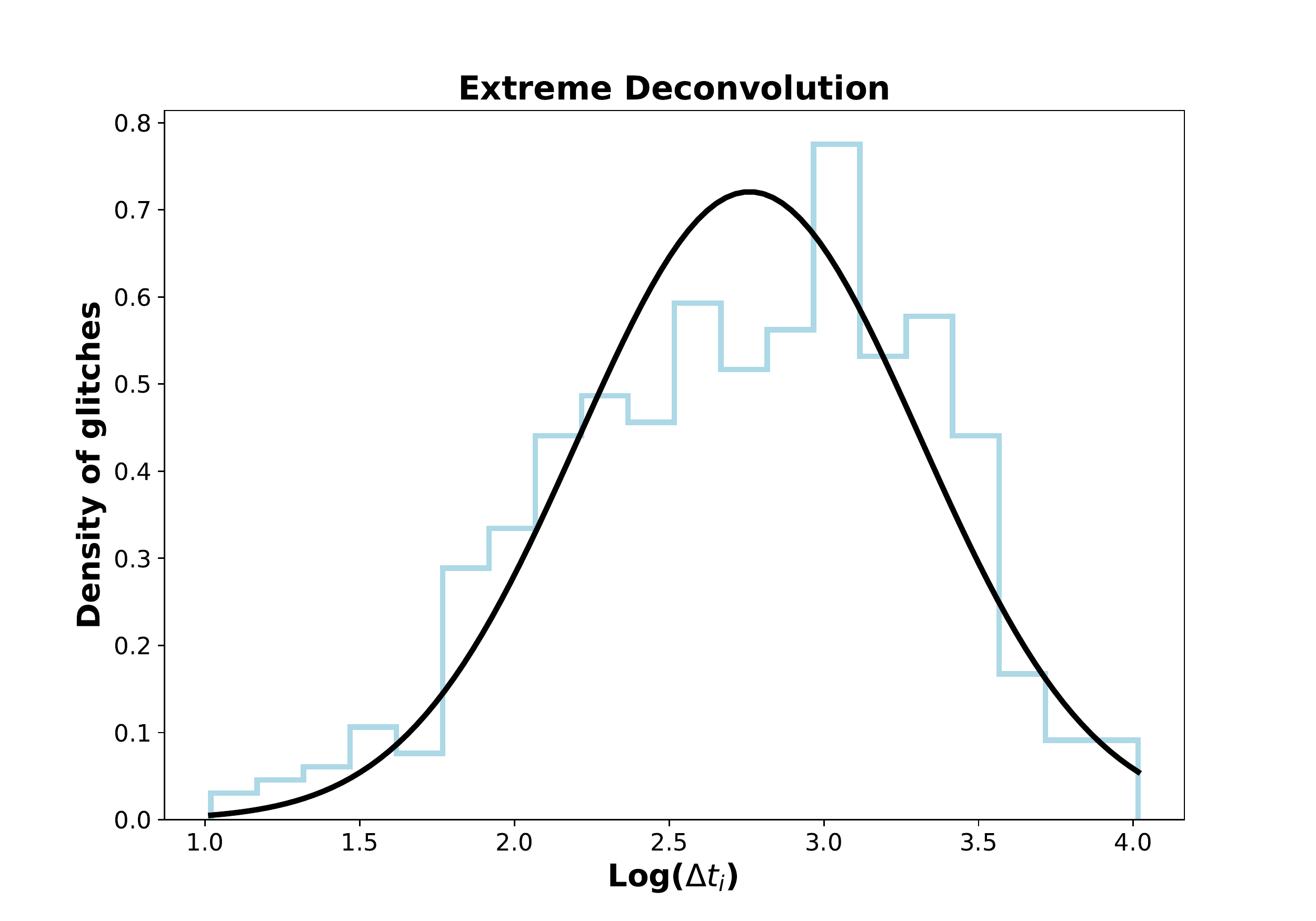}
\caption{Normalized histogram of the pulsar inter-glitch arrival time ($\log(\Delta t_i)$)(in days) along with the Gaussian fit obtained using XDGMM. The mean value and standard deviation for the Gaussian fit ($\mu,\sigma$) are given by (2.76, 0.55).}
\label{fig:interglitchhisto}
\end{figure}

\begin{figure}
\includegraphics[width=0.6\textwidth]{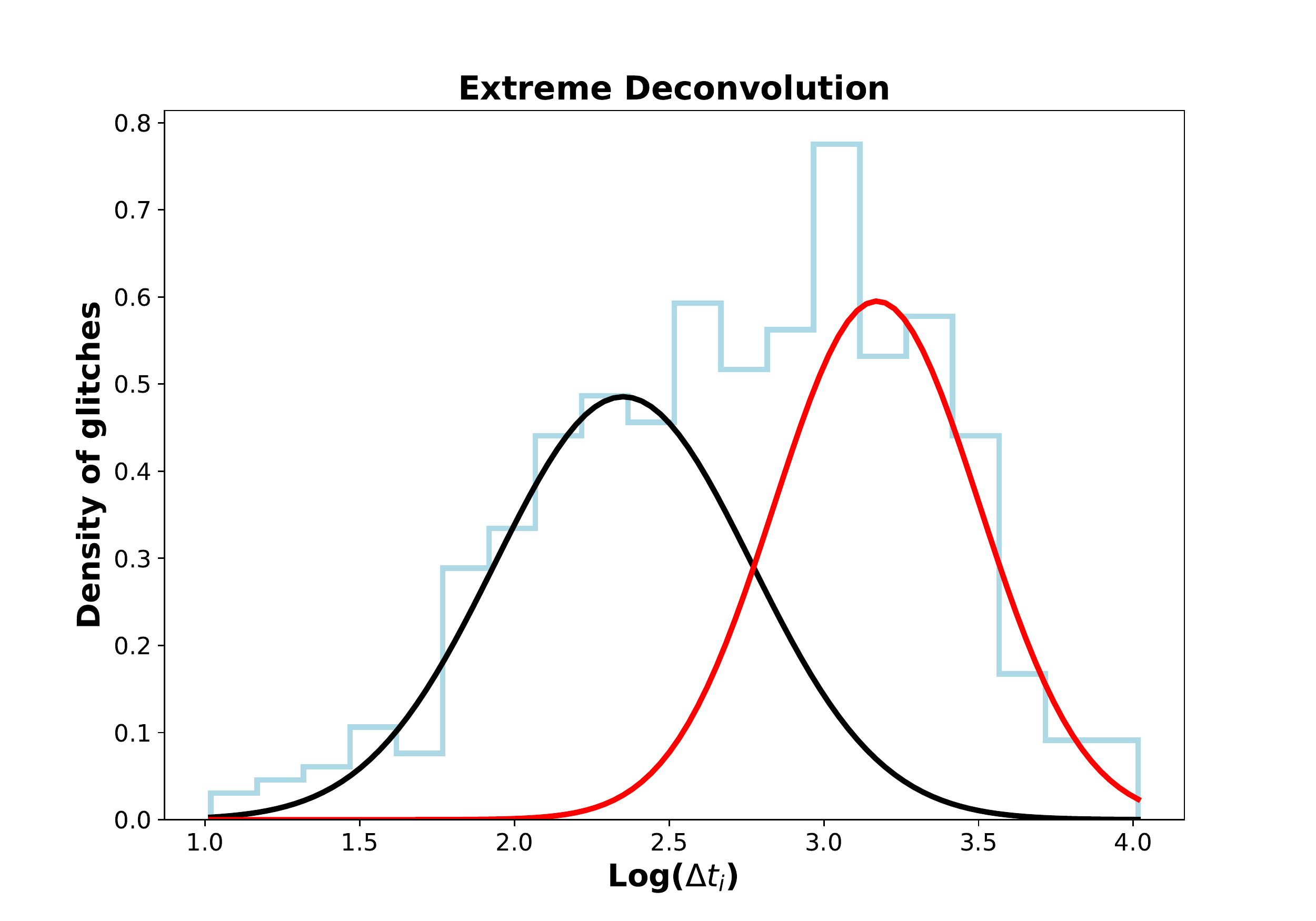}
\caption{Normalized histogram of the pulsar inter-glitch arrival time ($\log(\Delta t_i)$)(in days) along with the Gaussian fit obtained using XDGMM assuming two components. The mean value and standard deviation for the two components ($\mu,\sigma$) are given by (2.35, 0.41) and (3.17, 0.33) respectively.}
\label{fig:interglitchhisto2D}
\end{figure}

The results of XDGMM analysis can be found in Fig~\ref{fig:AIC2}. The histogram of the inter-glitch arrival times along with the XDGMM best-fit for one component is shown in Fig.~\ref{fig:interglitchhisto}, and the same with two components in Fig.~\ref{fig:interglitchhisto2D}. We find that the BIC analysis points  prefers an  unimodal distribution, whereas the AIC analysis shows a preference for two components with $\Delta$ BIC (AIC) between the first and second component equal to  -0.9 and 11.3, respectively. Therefore, AIC and BIC analyses lead to opposite conclusions.
We have seen this previously also where    AIC and BIC results did not always agree~\cite{Kulkarni17,Kulkarniexoplanet,Bhave}.  We note that AIC and BIC  answer different questions and are derived based on different assumptions. As emphasized by Tarnopolski~\cite{Tarnopolski19}, AIC chooses a model which describes reality for the data being analyzed, whereas BIC   finds the correct model among  the different hypotheses being tested. Therefore, they  differ in how they penalize the number of free parameters, where BIC is more stringent in penalizing parameters with additional free parameters~\cite{Tarnopolski19}. Therefore  BIC results are more trustworthy in case of a discrepancy, which is  seen here. Therefore, the current data for inter-glitch time interval is consistent with an unimodal distribution.

\section{Conclusions}
\label{sec:conclusions}
In this work, we use an automated method to determine the optimum number of radio pulsar glitch classes using the fractional glitch amplitude. Although a large number of works have previously alluded to a bimodal distribution for the pulsar glitch amplitude, very few works have implemented model selection techniques  to ascertain that two pulsar glitch components are decisively favored compared to other classes. Furthermore, no previous work has incorporated the observational uncertainties in their analysis.

To rectify this, we did a classification of \dnu  in logarithmic space, for all pulsar glitches collated in the JBO or ATNF catalog, using XDGMM which is an extension of the GMM technique, which incorporates the uncertainties in \dnu. We used information theory criteria such as AIC and BIC to determine the optimum glitch classes. Our results for the variation of AIC and BIC as a function of the number of the components can be found in Fig.~\ref{fig:AIC}. We see that both AIC and BIC have  a minimum value for  two components. The difference in BIC for two components is greater than 10 compared to any other value. Therefore, BIC decisively favors two components. The difference between the  AIC value for two and the next larger components is about five.

The dichotomy of the two glitch classes along with the XDGMM fit can be found in Fig.~\ref{fig:fermi2}. The mean values of \dnu for the two components are given by  $4.78 \times 10^{-9}$ and $1.38 \times 10^{-6}$, respectively. Therefore our results for the classification of fractional glitch amplitude are consistent with previous works~\cite{Lyne2000,Wang2000,Espinoza,Yu2012,Fuentes,Konar,Ashton,Eya2017,Eya2018,Basu2022}, which also pointed to a bimodal distribution, even though these works did not consider the uncertainty in the glitch amplitudes.

We have also applied this method to inter-glitch arrival time. The AIC/BIC distribution for this can be found in Fig.~\ref{fig:AIC2}. We find that BIC based analysis supports a single distribution, whereas  AIC analysis prefers two components.  The best-fit single and two component models for the inter-glitch intervals can be found in Fig.~\ref{fig:interglitchhisto} and Fig.~\ref{fig:interglitchhisto2D}. In such a scenario, the BIC based analysis is more trustworthy~\cite{Tarnopolski19}.

Therefore, our results on the autonomous classification of glitches corroborate previous results in literature which alluded to two distinct classes.  Since extensions to GMM can account for the observational uncertainties as well as incompleteness, it is very important to have an accurate characterization of not only all the  glitch observables, but also their uncertainties, completeness and upper limits,  as we head into the SKA era~\cite{Singha22}. We should obtain lot more statistics for pulsar glitches during the SKA era and incorporating the uncertainties and completeness into the analyses, could alter our  Physics inferences.  We can also apply our current technique to other pulsar glitch observables such as glitch activity, absolute glitch amplitude, etc. In order to carry out a similar analysis on the glitch activity, we would need an accurate estimate of the total time interval over which each pulsar has been monitored. Unfortunately,  this information is not publicly available at the time of writing. Once this has been made publicly available, we can apply the XDGMM technique for classification of the  glitch activity.

The unified data set used for this analysis along with the codes have been uploaded at  \url{https://github.com/swetha9730/XDGMM-for-Pulsar-Glitches-Classification}. 

\section*{Acknowledgements}
We are grateful to the anonymous referee for useful feedback and comments on our manuscript.
\bibliography{name}
\end{document}